\documentstyle[12pt]{article}
\begin{document}
\begin{center}
 {\LARGE Replica Symmetry Breaking in Renormalization:  Application to the
  Randomly Pinned Planar Flux Array}\\
\vspace{0.5cm}
 {\large Jan Kierfeld}\\
\vspace{0.25cm}
 {\large Institut f\"ur Theoretische Physik der Universit\"at zu K\"oln\\
         D-50923 K\"oln, Germany}
\end{center}

\noindent short title: Replica Symmetry Breaking in Renormalization
\begin{abstract}
The randomly pinned planar flux line array is supposed to show a phase
transition to a vortex glass phase at low temperatures. This transition
has    been  examined by using a  mapping  onto a 2D XY-model  with random
an\-iso\-tropy but without vortices and applying a renormalization group
treatment  to the replicated Hamiltonian based on the mapping
to a  Coulomb gas of vector charges. This renormalization group
approach is extended by deriving renormalization group flow equations
which take into  account  the possibility of  a one-step replica
symmetry breaking. It is shown that the renormalization group
 flow is unstable with respect to replica asymmetric perturbations
  and  new fixed points with a broken replica symmetry are obtained.
Approaching these fixed points the system can optimize its free energy
contributions from fluctuations on large length scales; an optimal
block size parameter $m$ can be found.
 Correlation functions for the case of a broken replica
symmetry can be calculated.  We obtain both  correlations
diverging as $\ln{r}$ and  $\ln^2{r}$ depending on the choice of $m$.
\end{abstract}
%
%
\section{Introduction}
The technological aspects of high-$\rm T_{c}$ superconductors
in strong magnetic fields and especially of their ability to preserve
superconductivity by flux pinning
\cite{Rev} have led to intense theoretical studies of the properties of a flux
line array in a type-II superconductor with random point-like pinning centers
\cite{F,NLS,TDV,K,GLD,TS,HwB,NKHw}.
It has been conjectured \cite{F,FGLV,N} that the flux
lines in a superconductor with point disorder form a new thermodynamic phase,
the vortex glass phase. It is supposed that in this phase the flux lines are
collectively pinned by the point defects and energy barriers between
different metastable states of the flux line array occur which diverge with
increasing length scale $L$ leading to a glassy dynamics and zero linear
resistivity \cite{TS,FFH,NKHw}. But there is still too little conclusive
evidence to confirm this scenario by analytic means.

Due to the absence of topological defects \cite{FFH,NKHw} in $1+1$ dimensions,
the planar $1+1$-dimensional flux line array can be well treated
analytically in an elastic approach. Actually, the system
of flux lines in a type-II superconducting plane with parallel magnetic
field and point disorder is the only system for which the existence of a
vortex glass phase has been shown analytically \cite{F,NLS,K,GLD,TS} by
applying various different methods.

On the other hand, the predictions for important physical features of
 the disordered $1+1$ dimensional
flux line array obtained by  the
different analytical methods
 differ significantly. In particular, there are still many
competing conjectures concerning the correlations in the vortex glass phase.
Moreover, the results of numerical simulations confirm neither of the
analytical predictions \cite{HwB}.
Essentially three analytic approaches have been applied to the problem:
(i) After using the replica trick and mapping onto a 2D XY-model with random
anisotropy but without vortices   \cite{F,NLS}, a renormalization
group (RG) calculation \cite{CO,GH}  has been  carried out with the replicated
Hamiltonian not  taking into account the possibility of replica
symmetry breaking (RSB). (ii) The replicated Hamiltonian has
also been studied by a variational treatment admitting of continuous RSB
 finding that  a one-step breaking is realized \cite{K,GLD}.
(iii) Without making use of the replica trick, the corresponding kinetic
equation has been  treated by a dynamical
 RG  analysis \cite{GS,TS}.

In the present paper we want to study how the concept of RSB could enter
into a RG analysis. For this purpose  we map the disordered
planar flux line array onto the 2D XY-model
with random anisotropy and perform a RG calculation with
the replicated Hamiltonian where we generalize the set of  coupling constants
such that we can take into account  a one-step RSB.
Our aim is  to show that an instability with
respect to one-step RSB can also be found in the RG treatment.
This leads to a more unified view of the approaches (i)
and (ii), and results obtained by the variational approach can partly be
reproduced in our calculation.

On scales larger than the flux line distance $l$ the planar flux line
array
is described by an elastic model with the positions of the flux lines
given by a scalar transverse displacement field $\phi({\bf r}) l/\pi$. The
field $\phi({\bf r})$ itself can be regarded as a phase field giving the
phase shift of the superconducting phase  caused by the flux line
displacements, because an increase in the flux line displacement by $l$
induces a shift of $\pi$ in the field $\phi({\bf r})$.
On large length scales,  the planar flux line
array interacting with the random pinning centers can then be described by the
Hamiltonian \cite{F,N,NLS,HwF}
\begin{equation} \label{1}
 \frac{1}{T}{\cal H}[\phi] = \int d^2{\bf r}
  \left\{ \frac{1}{2}K(\nabla\phi)^2
   + V_{1}({\bf r})\sin(2\phi({\bf r}))+V_{2}({\bf r})\cos(2\phi({\bf r}))
 \right\}.
\end{equation}
The second term containing the random potential ${\bf V}({\bf r})$ with
zero average and short range correlations ($i,j = 1,2$)
\begin{equation} \label{2}
  \overline{V_{i}({\bf r})V_{j}({\bf r'})} =  2 g~
  \delta_{\xi}({\bf r}-{\bf r'})\delta_{ij}
\end{equation}
models
the interaction of the flux lines with the point disorder in the continuum
description ($g$ includes a factor $1/T^2$).
The function $\delta_{\xi}$ is a delta-like function of the small width
$\xi$ given  by the maximum of the flux line core radius and the impurity
size.
A crucial feature of this term is to respect the periodicity
of the lattice, i.e.\ it is invariant under a uniform shift $\pi$ of the
field $\phi({\bf r})$. By rescaling of one coordinate the isotropic
first term for the elastic energy is obtained with one elastic constant
K (including a factor $1/T$).

Each of the three approaches sketched above \cite{F,NLS,CO,GH,K,GLD,GS,TS}
as well as the numerical
simulation \cite{HwB}, yield a phase transition at $K=1/\pi$ or $\tau=0$ with
\begin{equation} \label{3}
 \tau = 1-\frac{1}{\pi K} ~~,
\end{equation}
which serves as a small parameter measuring the distance from the
transition  and  controlling expansions around the transition.
For $\tau<0$ the system is in a high-temperature phase, disorder is not
relevant and does not alter the correlations induced by thermal
fluctuations on large length scales
$\overline{\langle (\phi({\bf r}) - \phi({\bf 0}))^2 \rangle} =
  (\ln{r})/(\pi K)$.

However, the results concerning the correlations for $\tau>0$ in the
glassy phase differ significantly. The RG analysis
carried out on the replica Hamiltonian in a replica symmetric way
yields correlations $\overline{\langle (\phi({\bf r}) -
 \phi({\bf 0}))^2 \rangle } \propto \tau^2 \ln^2{r}$ at  large length scales
$r$ \cite{CO,GH,TDV}.
The same result is obtained in the dynamical RG
calculation \cite{GS,TS}. Besides, correlations diverging like the square of
the logarithm follow from a real-space RG procedure \cite{VF}.
On the other hand, in the variational approach with one-step RSB,
correlations $\overline{\langle (\phi({\bf r}) - \phi({\bf 0}))^2 \rangle } =
 (\ln{r})/(\pi K (1-\tau)))$ are found \cite{K} to diverge logarithmically but
with a prefactor increasing with decreasing temperature.
Logarithmically diverging correlations
have also been found in the numerical simulation \cite{HwB}
but the prefactor of the logarithm does not accord with the
analytical prediction in \cite{K}.
Our calculation including one-step RSB in the RG analysis of the
replicated Hamiltonian  can reproduce both correlations diverging with
a simple logarithm and correlations diverging with the square of the logarithm
depending on the choice of the block size parameter $m$ in the one-step
RSB scheme.

\section{RG Analysis}
Introducing $n$ replicas and averaging over the disorder gives the
effective replicated Hamiltonian (with replica indices $\alpha$,$\beta$
running from 1 to n)
\begin{equation} \label{4}
 \frac{1}{T}{\cal H}_{R}[\phi_{\alpha}] = \int d^2{\bf r}
  \left\{ \frac{1}{2} \sum_{\alpha\beta} K_{\alpha\beta}
 \nabla\phi_{\alpha} \cdot \nabla\phi_{\beta} -
  \sum_{\alpha\beta} g_{\alpha\beta} \cos{(2(\phi_{\alpha}-\phi_{\beta}))}
 \right\}
\end{equation}
with matrices $K_{\alpha\beta}$ and $g_{\alpha\beta}$ taking on their
bare values
\begin{eqnarray}
 K_{\alpha\beta,0}&=&K \delta_{\alpha\beta}   \label{4a}\\
 g_{\alpha\beta,0}&=&g ~~. \label{4b}
\end{eqnarray}
 This Hamiltonian is equivalent to the
replica Hamiltonian of a 2D  XY-model with random
anisotropy but without vortices \cite{CO}.
The variational studies allowing for continuous RSB performed so far
on this Hamiltonian \cite{K,GLD} have shown that in the 2-dimensional model
considered here a one-step RSB is realized. Our calculation is restricted
so far to a one-step RSB scheme but the results from the variational
approach suggest that our results may stay valid even if it is possible
to extend the calculation to higher steps of RSB or a continuous RSB scheme.

Introducing one-step RSB and following the resulting RG flow, it is necessary
to admit  matrices $K_{\alpha\beta}$ and $g_{\alpha\beta}$ of the form
$K_{\alpha\beta} = A \delta_{\alpha\beta} +
B \widetilde{\delta}_{\alpha\beta} +C$ and $g_{\alpha\beta}= g_1
 \widetilde{\delta}_{\alpha\beta}+ g_2 (1-\widetilde{\delta}_{\alpha\beta})$;
 the
elements of the matrix $\widetilde{\delta}_{\alpha\beta}$ are 1 if $\alpha$
and $\beta$ belong to the same block of size $m$ and 0 otherwise.
Because of (\ref{4a}), (\ref{4b}) we have  initially
$A_0=K$, $B_0=C_0=0$ and $g_{1,0}=g_{2,0}=g$.
To derivate the full RG flow equations, we perform an analysis technically
similar to that of Cardy and Ostlund \cite{CO} but with significant
extensions to take into account the one-step RSB.
The calculation is based on the mapping onto a coupled Coulomb gas.
We want to consider only weak disorder, so initially the disorder strength
$g$ will be small;
 also throughout the RG procedure the matrix elements
of $g_{\alpha\beta}$ stay sufficiently small to use
 standard methods \cite{CO,JKKN} to transform the cos-couplings in the
partition sum and to integrate out
the fields $\phi_{\alpha}({\bf r})$
in favour of integer charges $n_{\alpha\beta}({\bf r})$  ($\alpha<\beta$)
with fugacity
$g_{\alpha\beta}$.
The replicated disorder averaged partition sum $\overline{Z^{n}}$ factors then
into $\overline{Z^{n}} = Z_{el}Z_{C}$ where the factor $Z_{el}$ represents
the purely elastic part of the partition sum; this factor plays a role
only in deriving the RG equation for the free energy density and will be
considered later on in detail.
In the Coulomb gas factor $Z_{C}$ of the  partition sum,
it has to be summed over all spatial configurations of interacting
charges, which can take on any integer value, but because the charge
fugacities $g_{\alpha\beta}$ are sufficiently small,  only positive  and
negative unit charges have to be considered, which obey  in
addition a neutrality condition.
Switching from the continuum description with a short wavelength cutoff
$\xi$ to a description on a square lattice with lattice constant $\xi$ and
lattice
vectors ${\bf R}$ for
easier notation, one obtains
 the following expression for the
replicated  disorder averaged partition sum:
\begin{eqnarray}
 \overline{Z^{n}} &=&Z_{el} Z_{C}~=~
  Z_{el} \times \prod_{\bf R} \prod_{\alpha<\beta}
 ~\sum_{n_{\alpha\beta}({\bf R})=-1}^{1} ~ \exp{(-{\cal H}_{C})} \label{5}\\
 -{\cal H}_{C}&=& \frac{1}{2} \sum_{{\bf R}\neq {\bf R'}}
    \sum_{\alpha,\beta,\gamma,\delta} n_{\alpha\beta}({\bf R}) ~
  (K^{-1})_{\beta\delta}~ n_{\gamma\delta}({\bf R'})~G'({\bf R}-{\bf R'})
   + \nonumber\\
  & &+ \sum_{\bf R} \sum_{\alpha\beta} ~(n_{\alpha\beta}({\bf R}))^2 ~
  \ln{g_{\alpha\beta}}~\label{6} ,
\end{eqnarray}
where $G'({\bf R}) = \ln{(|{\bf R}|/\xi)}$ and
 $n_{\beta\alpha}:=-n_{\alpha\beta}$ ($\alpha<\beta$). The matrix
$(K^{-1})_{\alpha\beta} = a \delta_{\alpha\beta} +
b \widetilde{\delta}_{\alpha\beta} +c$ has the same block
form as $K_{\alpha\beta}$
with $a=1/A$, $b=-B/A(A+mB)$ in the limit $n\to 0$.

 The block form of the matrix $g_{\alpha\beta}$
implies that two kinds of charges exist  differing in their fugacities
and, moreover, in their  interactions with other charges due to the block
form of the matrix $(K^{-1})_{\alpha\beta}$.

Taking this into account, a RG calculation
in the style of Cardy and Ostlund (CO) \cite{CO} can be performed, which
yields the following RG recursions in the limit $n\to 0$
upon a change of scale by a factor $e^l$ (Henceforth we always
include a factor $2/\pi$ in $a$,$b$ and $c$ and a factor
$4\pi\xi^2$ in $g_{1}$, $g_{2}$.):
\begin{eqnarray}
 da/dl&=& -{\textstyle \frac{1}{8}}~ a^2~ m~
   (g_{1}^2-g_{2}^2)    \label{flow1}  \\
 db/dl&=& {\textstyle \frac{1}{8}}~ a^2~
    (g_{1}^2-g_{2}^2 )  \label{flow2}   \\
 dc/dl &=& {\textstyle  \frac{1}{8}}~ a_{0}^2~
     g_{2}^2 \label{flow3}  \\
  dg_{1}/dl &=& (2- a) g_{1} ~-~
    {\textstyle \frac{1}{2}} (2-m) g_{1}^2
       ~-~ {\textstyle \frac{1}{2}} m g_{2}^2     \label{flow4}  \\
  dg_{2}/dl &=& (2-a-b) g_{2} ~-~ m g_{2}^2 ~-~
   (1-m)g_{1}g_{2}       \label{flow5}  ~~.
\end{eqnarray}
The parameter $m$ is a free parameter in these equations
with $0\le m\le 1$ in the limit $n\to 0$; possible choices for $m$ will
be discussed later on.
(\ref{flow1}), (\ref{flow2}) show that
\begin{equation} \label{11}
a+mb =a_{0} = 2-2\tau
\end{equation}
 is
not renormalized; this result is exact to all orders in the $g_{i}$
($i=1,2$) due to a
statistical invariance under tilt \cite{TDV,CO}.

In the special replica
symmetric cases $m=1$ and $m=0$, we get back the flow equations  of CO:
In the RG equations for $m=1$ ($m=0$), $g_{2}$ ($g_{1}$) plays the role
of the
single disorder strength parameter in \cite{CO},  the diagonal
matrix elements $a+b$ ($a$) of $K^{-1}_{\alpha\beta}$ are not renormalized,
and also the off-diagonal matrix elements
$c$ ($b+c$) and the fugacity $g_{2}$ ($g_{1}$) renormalize as in \cite{CO}.
The system exhibits the known
 CO fixed points
$g_{2}^{*}=0$ ($g_{1}^{*}=0$) and $g_{2}^{*}=2\tau$ ($g_{1}^{*}=2\tau$).
For $m=1$ the RG flow is sketched in Fig.~1a.
$g_{1}$ ($g_{2}$) does not feed back into the RG flow of the other
quantities and does  therefore  not enter  physical results like
correlation functions (see below).
 For this reason the introduction of a small initial
 replica asymmetric perturbation  ${\Delta g}_{0}=g_{1,0}-g_{2,0}$ has no
 effect on  physical results if $m=1$ ($m=0$),
 although ${\Delta g}$ turns out to be a relevant perturbation under RG
(see below).

Starting as in (\ref{4b})
 with replica symmetric initial conditions
$g_{1,0}=g_{2,0}$, replica symmetry is preserved throughout the RG
procedure independently of $m$,
 and $a$, $b$ are not renormalized; therefore the CO scenario with the
trivial fixed point $g_{1}^{*}=g_{2}^{*}=0$ and the non-trivial CO fixed point
$g_{1}^{*}=g_{2}^{*}=2\tau$
is  reproduced if replica symmetry holds initially.

However,  introducing a small initial replica asymmetry ${\Delta g}_{0}=
g_{1,0}-g_{2,0}\neq 0$ contrary to (\ref{4b}),
the RG flow develops for $\tau>0$
 an instability with respect to RSB.
The system flows for ${\Delta g}_{0} >0$ to a regime with $g_{1}>g_{2}$
and for ${\Delta g}_{0} <0$ to a regime $g_{1}<g_{2}$ (entering on large
length scales the unphysical regime of negative fugacities $g_{1}$). In
particular the replica symmetric CO fixed point $g_{1}^{*}=g_{2}^{*}
=2\tau$, $a^{*}=a_0$
is unstable against small replica asymmetric perturbations. A linear
stability analysis of the CO fixed point yields ($\Delta a=a_{0}-a$)
\begin{eqnarray}
  d\Delta a/dl &=& {\frac{1}{2}} m
   a_{0}^2\tau~ \Delta g    \label{ls1}  \\
  d\Delta g/dl &=& 2 {\frac{1}{m}}\tau
   ~{\Delta a}  \label{ls2} ~~.
\end{eqnarray}
These equations describe an instability with eigenvalue $2(1-\tau)\tau$
of the CO fixed point with respect to perturbations $\Delta g$. To avoid
entering the unphysical regime of negative fugacities, we consider only
perturbations ${\Delta g}_{0} >0$. As it is seen from (\ref{ls1}), (\ref{ls2}),
such a perturbation causes the charge interaction strength parameter $a$
 to decrease and  the asymmetry  $\Delta g$
to increase; finally, $a$ renormalizes to 0 following (\ref{flow1}).
This flow towards the fixed point $a^*=0$ implies that one non-interacting
 type of unit charge with fugacity $g_{1}$ appear
  on large length scales.
Furthermore, we can find from (\ref{flow4}), (\ref{flow5})
 two additional
  non-trivial RSB fixed points (\ref{13a}) and (\ref{13b}) with
$a^{*}=0$, $b^{*}=a_0/m$ for each of them:
\begin{eqnarray}
  g_{1}^{*} &=& 2 - {\frac{1-m}{m}}a_0
        + a_0 ~(1-  {\frac{2}{m}} + \frac{4}{a_0})^{1/2} ~~,
        \nonumber \\
 g_{2}^{*} &=& 2 - {\frac{2-m}{m}}a_0
        -  {\frac{1-m}{m}} ~
    a_0 ~(1-  {\frac{2}{m}} + \frac{4}{a_0})^{1/2} ~~~~
 \mbox{and}  \label{13a} \\
 g_{1}^{*} &=& {\frac{4}{2-m}} ~~,~~g_{2}^{*}~=~ 0 ~~.
    \label{13b}
\end{eqnarray}
At a certain
\begin{equation} \label{14}
   m^{*}=1-\tau/3+ {\cal O}(\tau^2)
\end{equation}
the fixed points (\ref{13a})
 and (\ref{13b}) fall exactly together.
Only for $m^{*}\le m<1$
the fixed point  (\ref{13a}) is in the physical regime
 $g_{2}^{*}\ge 0$ of non-negative fugacities. Moreover, the fixed point
 (\ref{13a}) is
 in this range of $m$ stable with respect to perturbations
in  $g_{1}$ and $g_{2}$ (getting marginal with respect to perturbations in
$g_{2}$ at $m=m^{*}$ where it coincides with (\ref{13b})),
whereas the fixed point
 (\ref{13b}) is  unstable with respect to perturbations $g_{2}>0$.
Therefore  the fixed point (\ref{13a}) is attractive for
all RG trajectories starting with $g_{1,0}>g_{2,0}>0$ (as illustrated in
Fig.~1b) while the fixed
point (\ref{13b}) is attractive only for RG  trajectories with
$g_{1,0}>g_{2,0}=0$.
For $0<m< m^{*}$ (\ref{13b}) is  the only RSB fixed point in the  physical
regime of non-negative fugacities $g_{2}\ge 0$. It is in this range of $m$ the
attractive fixed point for all RG trajectories with $g_{1,0}>g_{2,0}\ge 0$ (see
Fig.~1c,1d);
furthermore, it is stable with respect to perturbations in $g_{1}$ and
$g_{2}$.

As pointed out in (\ref{4a}), (\ref{4b}), the proper initial
values  $K_{\alpha\beta,0}=K \delta_{\alpha\beta}$ and
$g_{\alpha\beta,0}=g$ are replica symmetric with ${\Delta g}_{0}=0$.
It remains unclear in this approach how the initial asymmetry
${\Delta g}_{0}>0$ necessary for the development of an instability with
respect to RSB   can be obtained from physical reasons. One hint is
given in the next section where it is shown by comparison with the
replica symmetric CO  flow  for ${\Delta g}_{0}=0$  that contributions to the
free energy from large scale fluctuations can be optimized (which means
maximized in the limit $n\to 0$) if a small perturbation ${\Delta g}_{0}>0$
is introduced.

In the high-temperature phase for $\tau<0$ the system flows  to the stable
 trivial replica symmetric  fixed point $g_{1}^{*}=g_{2}^{*}=0$
 regardless of an initial
asymmetry ${\Delta g}_{0}\neq 0$. In this phase the trivial replica fixed
point is stable with respect to the RSB perturbation ${\Delta g}_{0}$
 so that RSB cannot occur
in the high-temperature phase as it is expected.
For $\tau=0$ the trivial fixed point stays marginally
stable.

\section{Free Energy and RSB}
We want  to proceed with a discussion of energetic aspects of the instability
in the RG flow upon introducing a replica asymmetric perturbation
 ${\Delta g}_{0} >0$  in the low-temperature phase.
 This enables us to fix the so far
undetermined block size parameter $m$ if  ${\Delta g}_{0} >0$ and to compare
the free energy in the RSB case with  ${\Delta g}_{0} >0$ with the free
energy in the replica symmetric CO case
 ${\Delta g}_{0} =0$.
The standard procedure to determine $m$ is to maximize
the free energy density per replica in the limit $n \to 0$
 with respect to the additional
free parameter $m$.

 In the RG approach it is possible to derive
the RG flow equation for the free energy density and calculate the
free energy by integrating the flow equation; moreover, one can separate
the contributions to the free energy from fluctuations on
 different length scales examining the flow of the free energy.
 In the RG procedure of
increasing the cutoff $\xi$ to $\xi e^{dl}$  in the partition sum and
rescaling the scale, one collects contributions not
renormalizing the coupling constants; these contributions
 enter the renormalization of the free
energy. Such terms are generated both in  the factor $Z_{C}$ by
contributions from integrating out the charge configurations
as described above  and in the factor
\begin{eqnarray}
 Z_{el} &=& \exp{\bigg( -{\frac{1}{2}} (L/2\pi\xi)^2
      \int_{-\pi}^{\pi}\int_{-\pi}^{\pi}
 d^2q  \bigg\{ n \ln{(\xi^2 q^2/2\pi)} +  } \nonumber\\
 & & ~~~~~~~~+ n [ {\frac{1-m}{m}}
   \ln{({\frac{\pi}{2}}a)} -  {\frac{1}{m}}
   \ln{({\frac{\pi}{2}}a_0)}
    +{\frac{\pi^2}{4}} c a_{0}] \bigg\} \bigg)
  \label{17}
\end{eqnarray}
(in the limit $n\to 0$ and with $L$ denoting the linear dimension  of the
system) by contributions from increasing the cutoff $\xi$ and adjusting
the couplings $a$, $c$ according to the flow equations (\ref{flow1}),
(\ref{flow3}).
Finally, one obtains in the limit $n \to 0$
 for the free energy density per replica
(apart from an additive constant independent of $m$)
the RG recursion relation
\begin{eqnarray} \label{18}
 {df/dl} &=& 2f + \frac{1}{n L^2} dZ/dl
 \nonumber\\
 &=&  2f ~-~ {\frac{1}{16\pi\xi^2}} (mg_{2}^2 + (1-m)g_{1}^2 )
   ~-~ \nonumber\\
 && ~-~ {\frac{1}{2\xi^2}}
  \bigg( [- {\frac{2(1-m)}{m}} \ln{(a_0/a)}
    -2  \ln{({\frac{\pi}{2}}a_0)}
  -{\frac{\pi^2}{2}}a_0 c] + \nonumber\\
 && ~~~~~~~~~ ~+~ [-{\frac{1-m}{m}}
     {\frac{1}{a}} {\frac{da}{dl}}] +
   [{\frac{\pi^2}{4}} a_0 {\frac{dc}{dl}}] \bigg)
  ~~.
\end{eqnarray}
{}From this recursion relation the initial free energy density $f_{0}$ can be
obtained by following the flow:
\begin{equation} \label{19}
 f_{0} ~=~ - \int_{0}^{\infty} dl e^{-2l}~
   {\frac{1}{n L^2} dZ/dl}  ~~.
\end{equation}
Using (\ref{18}), (\ref{19}) we want to study the contributions to the
 free energy from fluctuations on different length scales
 in the low-temperature phase  for  the
flow to the RSB fixed points (\ref{13a}), (\ref{13b}) induced by a
small initial perturbation ${\Delta g}_0>0$ and, in comparison,  for  the
 replica symmetric
flow to the CO fixed point starting with ${\Delta g}_0=0$.
To keep calculations tractable, we choose initial conditions in
the vicinity of the  CO fixed point, i.e.\ $g_{1,0}=2\tau+{\Delta g}_0$,
$g_{2,0}=2\tau$, when examining the RSB case.
For general initial conditions with ${\Delta g}_0>0$
there is initially a flow towards the CO fixed point slightly
perturbed by the small asymmetry  ${\Delta g}_0>0$; for the study of
the energetic effects of the RSB instability in the flow,
 this essentially replica symmetric part of the RG flow should be
negligible.

As a consequence of the factor $e^{-2l}$ appearing in the integrand of
(\ref{19}), the main contribution to $f_0$ comes from the short scales.
Starting at   $g_{1,0}=2\tau+{\Delta g}_0$,
$g_{2,0}=2\tau$ and evaluating (\ref{19}) straightforwardly to the
leading order in ${\Delta g}_0$, one obtains
a maximum of $f_0$ at the replica symmetric $m=0$. This is because the
main  contribution to  $-(dZ/dl)/nL^2$ comes on short scales from the
term $[-\pi^2 a_0 c/2]/2\xi^2$. The $m$-dependent part of
$[-\pi^2 a_0 c/2]/2\xi^2$  can be approximated by means
 of a linear stability analysis of the flow equations
(\ref{flow1})--(\ref{flow5}) at the CO fixed point
 enlarging on (\ref{ls1}), (\ref{ls2}) as
$[{\Delta g}_0 (\pi^2 a_0^3/16) (1-m)
 (\exp{(2\tau l)}-1)]/2\xi^2 $  with a maximum at
$m=0$.

 On the other
hand, the maximization of the asymptotic
 large scale contributions leads to a quite
different result.
 Examining the large scale contributions, one has to
investigate the asymptotics of the integrand in (\ref{19}) and to
maximize  $-(dZ/dl)/nL^2$  in the limit of $l \to \infty$.
For this purpose it is necessary to derive the asymptotics of $c$ which
is determined by the flow equation (\ref{flow3}).
{}From  (\ref{flow3}) follows that  for $m<m^{*} \le 1$ $c$ is asymptotically
linear divergent with an
asymptotics $c(l) \sim l~{g_{2}^{*}}^2 (1-\tau)^2/2$
where $g_{2}^{*}$ is taken in
the stable RSB fixed point (\ref{13a}), which is
$g_{2}^{*} \simeq  6(m-m^{*})$ to a good approximation.
However, we find in the regime  $0<m\le m^{*}$ from (\ref{flow3})
a saturation
of $c$ to a value $c^{*}$  because the stable RSB fixed point is in this
regime given by (\ref{13b}) with $g_{2}^{*}=0$.
To obtain an estimate for $c^{*}$ one has to determine
the characteristic scale $l^{*}$ on which $g_{2}$ renormalizes towards 0;
a linear stability analysis of   the flow
equations (\ref{flow1}), (\ref{flow4}), (\ref{flow5}) at the CO fixed point
 extending (\ref{ls1}), (\ref{ls2}) reveals that
$l^{*}$ can be approximated as
$l^{*} \simeq (\ln{(4\tau/(1-m){\Delta g}_0)})/2\tau $.
 From (\ref{flow3}) it  follows
 $c^{*}\simeq \tau\ln{(4\tau/(1-m){\Delta g}_0)}$
for the leading order contribution in $\tau$.
Moreover, it is seen
 from the
flow equation (\ref{flow1})  that $a(l) \sim 1/l$ on large scales.
Using these results for $c$ and $a$, one can verify easily from
(\ref{18}) that the most divergent
contributions to $-(dZ/dl)/nL^2$ come from
  $ [-2 \ln{(a_0/a) (1-m)/m} - \pi^2 a_0 c/2]/2\xi^2$
for large $l$. Maximization of these terms yields
\begin{equation} \label{20}
  m=m^{*} =1-\tau/3 + {\cal O}(\tau^2) ~~,
\end{equation}
because the maximization of the second term restricts $m$ to values
$0<m\le m^{*}$ to avoid the occurrence of the linear divergence in the
regime $m^{*}<m\le 1$ and the maximization of the only
 logarithmically diverging first term singles out the greatest value
$m=m^{*}$ of the interval $0<m\le m^{*}$. (\ref{20}) is in fairly good
agreement with \cite{K}.

RSB is a large scale effect associated
with the existence of diverging energy barriers generating metastable
states. Therefore it seems to be more reasonable to consider only
the large scale contributions to the free energy in (\ref{19})
although  the expression (\ref{19}) for $f_0$ is dominated by its short
scale part.  This
is equivalent to considering the free energy of the renormalized but
not rescaled Hamiltonian on large scales but discarding a constant
  energy shift
depending on $m$ which comes from short scales. This energy shift,
which is essentially replica symmetric, may describe the
free energy of  single metastable states.
 In the presence of an initial
asymmetric perturbation  ${\Delta g}_{0} >0$,  maximization of the large
scale contributions to the free energy yields then   a maximum at $m=m^{*}$
as derivated above.

Comparison of  these large scale contributions for the flow to the RSB
 fixed point
when ( ${\Delta g}_{0} >0$) and for the replica symmetric flow to the
CO fixed point ( ${\Delta g}_{0} =0$) shows that this part of the free energy
is greater in the RSB case. This is because the most divergent contribution
  to  $-(dZ/dl)/nL^2$  is in the replica symmetric case as in the RSB case
 given by  $[-\pi^2 a_0 c/2]/2\xi^2$ with
$c(l) \sim l~ {g_{2}^{*}}^2 (1-\tau)^2/2$
but the replica symmetric CO fixed
 point value $2\tau$ for  $g_{2}^{*}$ is always greater than
 or equal to ($m=1$) the RSB fixed
point values given by  (\ref{13a}), (\ref{13b}).
Therefore it is energetically favorable on large scales to break
the replica symmetry by introducing a perturbation  ${\Delta g}_{0} >0$.
This energy gain can occur  on scales larger than
$L^{*}=\xi \exp{(l^{*}|_{m=m^{*}})}=
\xi(4\tau/(1-m^{*}){\Delta g}_0)^{1/2\tau}=\xi (12/{\Delta g}_0)^{1/2\tau}$.

In the high-temperature phase the trivial  fixed point    is stable
with respect to the introduction of a  perturbation  ${\Delta g}_{0}\neq 0$.
For this reason the large scale contributions to the free energy are the same
as in the replica symmetric case. For the short scale contributions
to the free energy  the same argumentation applies as in the low-temperature
phase leading to a maximum at the
replica symmetric $m=1$ if ${\Delta g}_{0}\neq 0$. So
RSB is energetically not favorable in the high-temperature phase as
it is expected.

\section{Correlations}
The  RG flow and fixed point structure changes  significantly
upon introducing an energetically favorable,
 replica asymmetric perturbation  ${\Delta g}_{0} >0$
in the low-tem\-per\-at\-ure
phase as outlined above.
As well
the behaviour of the $\overline{\langle\phi\phi\rangle}$-correlations
changes drastically
depending on the value of $m$. The Fourier transformed  correlations
 between replicas on large scales  can be calculated  by using a
Gaussian approximation to the renormalized but not rescaled replica
Hamiltonian, which yields for small $q\sim e^{-l}/\xi$ \cite{TDV}
\begin{equation} \label{15}
 \langle \phi_{\alpha}({\bf q}) \phi_{\beta}(-{\bf q}) \rangle
   ~=~ \frac{1}{q^2}~ (K^{-1})_{\alpha\beta}(l=\ln{(\xi/q)}) ~~,
\end{equation}
so that the large scale correlations depend essentially only on the
asymptotic RG flow of the matrix elements $a$, $b$ and $c$.

 From expression
(\ref{15}) one can verify that the connected correlation function
\begin{equation} \label{15a}
\overline{\langle\phi({\bf q})\phi(-{\bf q})\rangle -
\langle\phi({\bf q})\rangle\langle\phi(-{\bf q})\rangle}=
\lim_{n\to 0} 1/n \sum_{\alpha,\beta}  \langle \phi_{\alpha}({\bf q})
 \phi_{\beta}(-{\bf q}) \rangle= (1-\tau)\pi/q^2
\end{equation}
 does not change its form at the transition, independently  from the
introduction of a nonzero ${\Delta g}_{0}$
  due to the non-renormalization of $a+mb=a_0$, contrary to
the $\overline{\langle\phi\phi\rangle}$-correlations \cite{K}.

For the $\overline{\langle\phi\phi\rangle}$-correlations (\ref{15}) yields
\begin{equation} \label{16}
 \overline{\langle\phi({\bf q})\phi(-{\bf q})\rangle} ~=~
  \langle \phi_{\alpha}({\bf q})\phi_{\alpha}(-{\bf q}) \rangle ~=~
  \frac{\pi}{2q^2}~ (a(l)+b(l)+c(l))|_{l=\ln{(\xi/q)}} ~~.
\end{equation}
In the high-temperature phase no RSB takes place,  even if ${\Delta g}_{0}>0$,
 and
there is essentially no renormalization of $a$, $b$ and $c$ ; the
connected $\overline{\langle\phi\phi\rangle-
\langle\phi\rangle\langle\phi\rangle}$-correlation function and the
$\overline{\langle\phi\phi\rangle}$-correlation function coincide then
and $\overline{\langle (\phi({\bf r}) - \phi({\bf 0}))^2 \rangle}
= (1-\tau) \ln{(r/\xi)}$.

In the low-temperature phase the asymptotics of $c$, which is determined by
the flow equation (\ref{flow3}), is
of special interest  because in the
replica symmetric case, i.e.\  without a replica asymmetric perturbation
(${\Delta g}_0 =0$),  the asymptotics
$c \sim 2(1-\tau)^2\tau^2 ~l$ diverging linearly  gives
correlations $\overline{\langle (\phi({\bf r}) - \phi({\bf 0}))^2 \rangle}
\sim  ((1-\tau)^2\tau^2/2) \ln^2{(r/\xi)}$
 diverging  with $\ln^2{r}$ \cite{TDV}.

In the RSB case with an initial ${\Delta g}_0 >0$,
$c$ has  also a linear divergent  asymptotics
$c(l) \sim l~ {g_{2}^{*}}^2 (1-\tau)^2/2$ for
$m^{*}<m\le 1$ (see above), where
$g_{2}^{*} \simeq  6(m-m^{*})$  is taken in
the stable RSB fixed point (\ref{13a}). This  entails
 $\overline{\langle\phi\phi\rangle}$-correlations diverging also with
 $\ln^2{r}$ for this range of $m$ but with a prefactor reduced by a factor
$9(m-m^{*})^2/\tau^2<1$ compared to the replica symmetric case;
in particular, we get back the replica symmetric CO result choosing $m=1$.
 The
situation changes significantly in the regime $0<m\le m^{*}$ because the
stable RSB fixed point is in this regime given by
 (\ref{13b}) with $g_{2}^{*}=0$.
Therefore $c$ saturates on large scales to a value
 $c^{*}\simeq \tau\ln{(4\tau/(1-m){\Delta g}_0)} $
for the leading order contribution in $\tau$ (see above)
 and we obtain from (\ref{16}) only logarithmically divergent
$\overline{\langle\phi\phi\rangle}$-correlations with a
prefactor $a_0/2m+c^{*}/2$ which is  greater than in the
high-temperature phase.
To the leading order in $\tau$ this yields  correlations
$\overline{\langle (\phi({\bf r}) - \phi({\bf 0}))^2 \rangle}
\simeq((1-\tau)/m ~+~\tau(\ln{(4\tau/(1-m){\Delta g}_0)})/2 )  \ln{(r/\xi)}$;
with our above choice (\ref{20}) of $m \simeq 1-\tau/3$  we get
$\overline{\langle (\phi({\bf r}) - \phi({\bf 0}))^2 \rangle}
\simeq((1-2\tau/3) +\tau(\ln{(12/{\Delta g}_0)})/2 ) \ln{(r/\xi)}$.
This implies that the  prefactor of the logarithm
 increases with $\tau$  in the low-temperature phase.

Our results for the low-temperature phase show that within the
one-step RSB RG approach  with  a small initial replica
asymmetry  ${\Delta g}_0 >0$,  it is possible to obtain the known
replica symmetric result for the
 $\overline{\langle\phi\phi\rangle}$-correlations   diverging
like $\ln^2{r}$ \cite{TDV} if $m=1$  as well as
 $\overline{\langle\phi\phi\rangle}$-correlations diverging in the same way
but with a smaller prefactor if $m^{*}<m \le 1$  and logarithmically
divergent  $\overline{\langle\phi\phi\rangle}$-correlations with
a prefactor increasing with decreasing temperature if $0<m \le m^{*}$.
The latter possibilities are of interest with regard to the numerical
simulations \cite{HwB} and the results of the variational approaches
\cite{K,GLD}.

\section{Conclusion}

To summarize we have shown an instability in the RG flow
 of the disordered planar flux
line array, which is equivalent to the 2D XY-model with random anisotropy but
without vortices, with respect to a one-step RSB. The flow approaches
 new RSB fixed points
if an initial replica asymmetric perturbation is introduced.
The system can optimize its free energy contributions from large length
scale ($>L^{*}$) fluctuations
 by breaking the replica symmetry and approaching the
RSB fixed point where the energetical optimal choice of $m$ is
$m=m^{*}\simeq 1-\tau/3$.
Introducing the initial  perturbation,
  the  $\overline{\langle\phi\phi\rangle}$-correlations show
a $\ln^2{r}$-divergence on large length scales in the range
$m^{*}<m \le 1$ returning to the replica symmetric result at $m=1$;
for $0<m \le m^{*}$ the correlations diverge only as $\ln{r}$, which
is especially for $m=m^{*}$ the case.

During completion of this work P. Le Doussal and T. Giamarchi have
submitted a letter \cite{GLD2} in which they find independently from
our results an instability with respect to RSB in the 2D XY model
in a random field.

\section{Acknowledgements}
The author thanks T. Nattermann, S.E. Korshunov
 and T. Hwa for discussions and SFB 341 (B8)
for support.

\vspace{2cm}
\noindent {\bf Figures Caption}

\vspace{0.3cm}
\noindent
\underline{Fig. 1:} RG flow trajectories for $\tau=0.05$ and different
values of $m$ with an
initial replica asymmetry ${\Delta g}_{0}=g_{1,0}-g_{2,0}=\tau^2$;
a) $m=1$, b) $m=(1+m^{*})/2$, c) $m=m^{*}$, d) $m=1-5\tau$.
The dashed line is the line $g_{1}=g_{2}$ of replica symmetric values. The
trivial fixed point and the CO fixed point on this line
 as well as the RSB fixed
points (\ref{13a}), (\ref{13b}) are plotted.


\begin{thebibliography}{99}
\bibitem{Rev}
G. Blatter {\em et al}, Rev. Mod. Phys. (to appear)
\bibitem{F}
M.P.A. Fisher, Phys. Rev. Lett. {\bf 62}, 1415 (1989).
\bibitem{FGLV}
M.V. Feigelman, V.B. Geshkenbein, A.I. Larkin and V.M. Vinokur,
Phys. Rev. Lett. {\bf 63}, 2303 (1989).
\bibitem{N}
T. Nattermann, Phys. Rev. Lett. {\bf 64}, 2454 (1990).
\bibitem{NLS}
T. Nattermann, I. Lyuksyutov and M. Schwartz, Europhys. Lett. {\bf 16}, 295
(1991).
\bibitem{CO}
J.L. Cardy and S. Ostlund, Phys. Rev. B {\bf 25}, 6899 (1982).
\bibitem{GH}
Y.Y. Goldschmidt and A. Houghton, Nucl. Phys. {\bf B210}, 115 (1982).
\bibitem{TDV}
J. Toner and D.P. DiVincenzo, Phys. Rev. B {\bf 41}, 632 (1990).
\bibitem{K}
S.E. Korshunov, Phys. Rev. B {\bf 48}, 3969 (1993).
\bibitem{GLD}
T. Giamarchi and P. Le Doussal, Phys. Rev. Lett. {\bf 72}, 1530 (1994).
\bibitem{GS}
Y.Y. Goldschmidt and B. Schaub, Nucl. Phys. {\bf B251}, 77 (1985).
\bibitem{TS}
Y.C. Tsai and Y. Shapir, Phys. Rev. Lett. {\bf 69}, 1773 (1992) and
       {\bf 71}, 2348 (1993).
\bibitem{VF}
J. Villain and J.F. Fernandez, Z. Phys. B {\bf 54}, 139 (1984).
\bibitem{HwB}
G.G. Batrouni and T. Hwa, Phys. Rev. Lett. {\bf 72}, 4133 (1994).
\bibitem{FFH}
D.S. Fisher, M.P.A. Fisher and D.A. Huse, Phys. Rev. B {\bf 43}, 130 (1991).
\bibitem{NKHw}
T. Nattermann, J. Kierfeld and T. Hwa, to be published.
\bibitem{HwF}
T. Hwa and D.S. Fisher, Phys. Rev. Lett. {\bf 72}, 2466 (1994).
\bibitem {JKKN}
 J. Jos\'{e}, L.P. Kadanoff, S. Kirkpatrick und
               D.R. Nelson, Phys. Rev. B {\bf 16}, 1217 (1977)
\bibitem{GLD2}
P. Le Doussal and T. Giamarchi, submitted to Phys. Rev. Lett.
\end{thebibliography}
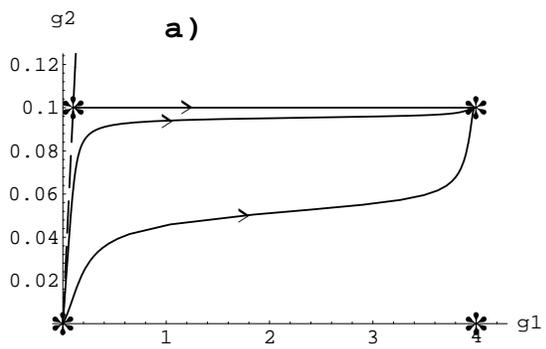
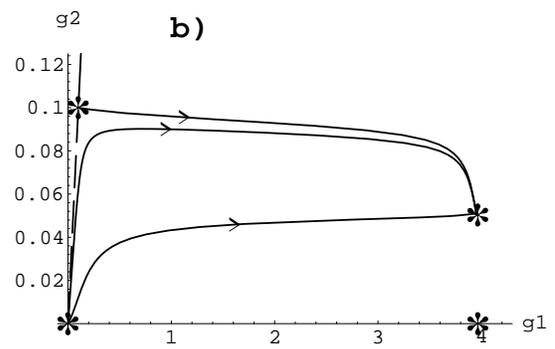
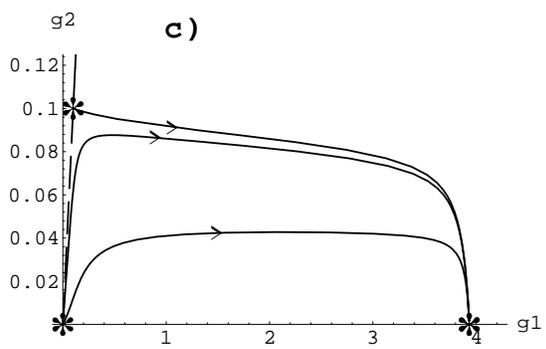
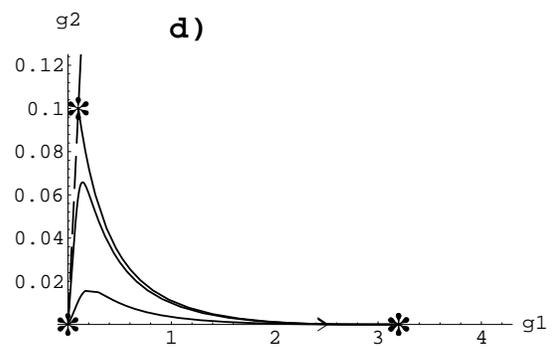
\end{document}